\documentclass[11pt]{llncs}
\usepackage{amsmath}

\usepackage{color}
\usepackage{times, amssymb}

\newcommand{\defeq}{\stackrel{\Delta}{=}}
\newcommand{\ket}[1]{| #1 \rangle}

\newcommand{\braket}[2]{\langle #1 | #2 \rangle}

\newcommand{\hy}{\hat{y}}

\newcommand{\amp}{\mathsf{A}}

\newcommand{\tensor}{\otimes}

\begin{document}

\title{Is partial quantum search of a database any easier?}
\author{ Lov K.\ Grover\thanks{%
Research was partly supported by NSA \& ARO under contract
no.~DAAG55-98-C-0040.} \inst{1} \and Jaikumar Radhakrishnan\thanks{%
This work was done during the author's visit to Bell Labs which was
supported by the above contract; current affiliation: 
Toyota Technological Institute at Chicago, 1427 E. 60th Street, Chicago, IL
60637.
}\inst{2} }

\authorrunning{Grover and Radhakrishnan}

\tocauthor{Lov K.\ Grover (Bell Labs) and
Jaikumar Radhakrishnan (Tata Institute of Fundamental Research)
}

\institute{
Bell Laboratories,
Lucent Technologies,\\
600--700 Mountain Avenue, Murray Hill, NJ 07974\\
e-mail: lkgrover@bell-labs.com
\and
School of Technology and Computer Science,\\
Tata Institute of Fundamental Research, Mumbai 400005, India\\
e-mail:jaikumar@tifr.res.in
}

\maketitle

\begin{abstract}
We consider the partial database search problem where given a quantum
database $\{0,1\}^{n}\overset{f}{\rightarrow}\{0,1\}$ with the condition that
$f(x) =1$ for a unique $x \in \{0,1\}^n$, we are required to determine
\textit{only} the first $k$ bits of the address $x$. We derive an
algorithm and a lower bound for this problem. Let $q(k,n)$ be the
minimum number of queries needed to find the first $k$ bits of the
required address $x$ with certainty (or with very
high probability, say $1-O(N^{-\frac{1}{4}})$). We show that there
exist constants $c_{k}$ (corresponding to the algorithm) and $d_{k}$
(corresponding to the lower bound) such that
\[
\frac{\pi }{4}\left( 1-\frac{d_{k}}{\sqrt{K}}\right) \sqrt{N}\leq q(k,n)\leq 
\frac{\pi }{4}\left( 1-\frac{c_{k}}{\sqrt{K}}\right) \sqrt{N},
\]%
where $K=2^{k}$ and $N=2^{n}$.  Our algorithm returns the correct
answer with probability $1-O(1/\sqrt{N})$, and can be easily modified
to give the correct answer with certainty. The lower bound for
algorithms that return the correct answer with certainty is proved by
reducing the usual database search problem to this partial search
problem, and invoking Zalka's lower bound showing that Grovers
algorithm is optimal for the usual database search problem.  We then
derive a lower bound that is applicable for database search algorithms
that err with small probability, and use it to show that our lower
bound also applies to partial search algorithms that return the
correct answer with probability at least
$1-O(N^{-\frac{1}{4}})$. Thus, it is easier to determine some $k$ bits
of the target address than to find the entire address, but as $k$
becomes large this advantage reduces rapidly.

%
%
%

\end{abstract}

\authorrunning{Grover and Radhakrishnan}

\tocauthor{Lov K.\ Grover (Bell Labs) and
Jaikumar Radhakrishnan (Tata Institute of Fundamental Research)
}

\institute{
Bell Laboratories,
Lucent Technologies,\\
600--700 Mountain Avenue, Murray Hill, NJ 07974\\
e-mail: lkgrover@bell-labs.com
\and
School of Technology and Computer Science,\\
Tata Institute of Fundamental Research, Mumbai 400005, India\\
e-mail:jaikumar@tifr.res.in
}

\newpage

\section{Introduction}

%
%
In the database search problem, we are required to determine the
address where a given item is stored. In this paper, we consider a
variant of the problem, where we do not want the entire address but
only its first $k$ bits.  The motivation for considering this problem
is that quite often in real databases, just a small part of the target
address contains all the information we need. For example: the items
in a database may be listed according to the order of preference (say
a merit-list which consists of a ranking of students in a class sorted
by the rank). We want to know roughly where a particular student
stands---we might want to know whether he/she ranks in the top 25\%,
the next 25\%, the next 25\%, or the bottom 25\%. In other words, we
want to know the first two bits of the rank and since the list was
sorted by rank, we need just the first two bits of the address.

\subsection{Partial classical search is easier}

First consider classical algorithms. We will assume that our
algorithms make no errors. Using a simple randomized classical search
algorithm one can find an element in such a database using, on an
average, $\frac{N}{2}$ queries ($N$ is the number of elements in the
database). This bound is tight if we wish to to determine the location
of the element exactly.

Now, consider partial search. That is, the range $\{1,2,\ldots ,N\}$
into $K$ equal blocks (in the example above, these blocks are
intervals) and we are asked to determine in which of these blocks the
target item lies. A natural algorithm for this problem is to randomly
choose $K-1$ of the $K$ blocks and probe the locations in these blocks
in some random order.  If the target lies in the block that is being
searched, the algorithm will find it; however, if it does not find the
target, we come to know that it is one of the $\frac{N}{K}$ items that
the algorithm has not probed. A simple analysis shows that the
expected number of queries made by this algorithm is
$\frac{N}{2}\left( 1-\frac{1}{K^{2}}\right)$. So, this algorithm saves
a little for the seemingly easier problem. It can be verified that no
classical randomized algorithm can do better (see
Appendix~\ref{app:randomized} for a proof). So, the savings
necessarily reduce very fast as $K$ grows. If we restrict ourselves to
deterministic algorithms, then using the same idea we can derive an
algorithm that makes only $N\left( 1-\frac{1}{K}\right) $ queries, a
saving of $\frac{N}{K}$ queries over any algorithm that solves the
original database problem with certainty.

\subsection{Why partial quantum search may be easier}

In this paper we study this problem in the quantum setting.  Again, we
restrict ourselves to algorithms that make no error.  It is well known
that if the database is supplied in the form of a suitable quantum
oracle, then it is possible to use \emph{quantum parallelism} and
determine the rank of any element using $\left( \frac{\pi }{4}\right)
\sqrt{N}$ queries~\cite{Grover}. Furthermore, this algorithm is
optimal~\cite{Zalka} (also \cite{Bennett,Boyer}).  The ideas used to
speed up partial search classically can be used to reduce the number
of queries by a factor of $\sqrt{\frac{K-1}{K}}\approx 1-\frac{1}{2K}$
over the standard quantum search algorithm. That is, we randomly pick
$K-1$ of the blocks and run the quantum search algorithm on the
$N\left(1-\frac{1}{K}\right)$ locations in the chosen blocks.  This
would require
\[
\frac{\pi }{4}\sqrt{\frac{(K-1)N}{K}}~\approx ~\frac{\pi }{4}\left(1-\frac{1}{2K}\right)
\sqrt{N}
\]%
queries. As in the deterministic classical search algorithm, the
savings in the number of queries are $O(\frac{1}{K})$ times
the number of queries needed for exact search.  In the classical case,
that is the best that we could do. In this paper, we show a better
quantum algorithm that saves 
$\theta(\frac{1}{\sqrt{K}})$ times the number number of queries needed
for exact search, and argue that up to constant factors this the best
one can hope for. Our main result is the following.
\begin{theorem}
Assume that $N$ is much larger
than $K$.
\begin{description}
\item[Upper bound:] For all $K$ there is a constant $c_k>0$, such that
there is a quantum algorithm for partial search that makes at most
$(\frac{\pi}{4})(1-c_K) \sqrt{N}$ queries to the database and returns
the answer with probability at least $1-O(N^{-\frac{1}{2}})$; this
algorithm can be modified to return the correct answer with certainty
while increasing the number of queries by at most a constant. For
large $K$, we have $c_K \geq \frac{0.42}{\sqrt{K}}$.

\item[Lower bound:] Fix $K$. Suppose there is a quantum algorithm for
partial search that makes at most $\frac{\pi}{4} (1-d_k) \sqrt{N}$
queries (for all large $N$) and returns the correct answer with
probability at least $1-O(N^{-\frac{1}{4}})$, then $d_K \leq
\frac{1}{\sqrt{K}}$.
\end{description}
\end{theorem}

The upper bound is obtained by judicious combinations of amplitude
amplification steps used for the standard quantum database search
problem. The lower bound is obtained by observing that an algorithm
for the regular database search problem can be obtained by using the
partial search algorithm as a subroutine and using lower bounds
for the regular database search (for the zero-error case we can
use Zalka~\cite{Zalka}'s result, which we refine in order to derive
our lower bounds for partial search algorithms with small error).

\subsection{An example}

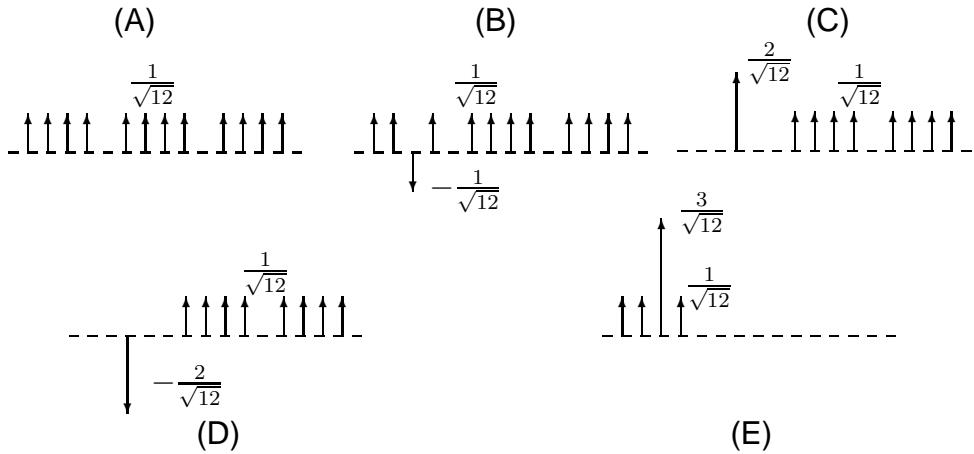
\newcommand{\amplone}{$\frac{1}{\sqrt{12}}$}
\newcommand{\ampltwo}{\amplone}
\newcommand{\amplsix}{\amplone}
\newcommand{\amplthree}{$-\frac{1}{\sqrt{12}}$}
\newcommand{\amplfour}{$\frac{2}{\sqrt{12}}$}
\newcommand{\amplfive}{$-\frac{2}{\sqrt{12}}$}
\newcommand{\amplseven}{$\frac{3}{\sqrt{12}}$}
\begin{figure}[thb]
\begin{center}
\setlength{\unitlength}{4144sp}%
\begingroup\makeatletter\ifx\SetFigFont\undefined
\def\x#1#2#3#4#5#6#7\relax{\def\x{#1#2#3#4#5#6}}%
\expandafter\x\fmtname xxxxxx\relax \def\y{splain}%
\ifx\x\y   
\gdef\SetFigFont#1#2#3{%
  \ifnum #1<17\tiny\else \ifnum #1<20\small\else
  \ifnum #1<24\normalsize\else \ifnum #1<29\large\else
  \ifnum #1<34\Large\else \ifnum #1<41\LARGE\else
     \huge\fi\fi\fi\fi\fi\fi
  \csname #3\endcsname}%
\else
\gdef\SetFigFont#1#2#3{\begingroup
  \count@#1\relax \ifnum 25<\count@\count@25\fi
  \def\x{\endgroup\@setsize\SetFigFont{#2pt}}%
  \expandafter\x
    \csname \romannumeral\the\count@ pt\expandafter\endcsname
    \csname @\romannumeral\the\count@ pt\endcsname
  \csname #3\endcsname}%
\fi
\fi\endgroup
\begin{picture}(5780,2673)(261,-2770)
\thinlines
{\color[rgb]{0,0,0}\put(390,-970){\vector( 0, 1){234}}
}%
{\color[rgb]{0,0,0}\put(507,-970){\vector( 0, 1){234}}
}%
{\color[rgb]{0,0,0}\put(623,-970){\vector( 0, 1){234}}
}%
{\color[rgb]{0,0,0}\put(740,-970){\vector( 0, 1){234}}
}%
{\color[rgb]{0,0,0}\put(974,-970){\vector( 0, 1){234}}
}%
{\color[rgb]{0,0,0}\put(1091,-970){\vector( 0, 1){234}}
}%
{\color[rgb]{0,0,0}\put(1208,-970){\vector( 0, 1){234}}
}%
{\color[rgb]{0,0,0}\put(1325,-970){\vector( 0, 1){234}}
}%
{\color[rgb]{0,0,0}\put(1558,-970){\vector( 0, 1){234}}
}%
{\color[rgb]{0,0,0}\put(1675,-970){\vector( 0, 1){234}}
}%
{\color[rgb]{0,0,0}\put(1792,-970){\vector( 0, 1){234}}
}%
{\color[rgb]{0,0,0}\put(1909,-970){\vector( 0, 1){234}}
}%
{\color[rgb]{0,0,0}\multiput(273,-970)(113.09677,0.00000){16}{\line( 1, 0){ 56.548}}
}%
{\color[rgb]{0,0,0}\multiput(3826,-2068)(113.09677,0.00000){16}{\line( 1, 0){ 56.548}}
}%
{\color[rgb]{0,0,0}\put(3943,-2068){\vector( 0, 1){234}}
}%
{\color[rgb]{0,0,0}\put(4060,-2068){\vector( 0, 1){234}}
}%
{\color[rgb]{0,0,0}\put(4293,-2068){\vector( 0, 1){234}}
}%
{\color[rgb]{0,0,0}\put(4176,-2068){\vector( 0, 1){702}}
}%
{\color[rgb]{0,0,0}\put(1334,-2068){\vector( 0, 1){233}}
}%
{\color[rgb]{0,0,0}\put(1451,-2068){\vector( 0, 1){233}}
}%
{\color[rgb]{0,0,0}\put(1568,-2068){\vector( 0, 1){233}}
}%
{\color[rgb]{0,0,0}\put(1685,-2068){\vector( 0, 1){233}}
}%
{\color[rgb]{0,0,0}\put(1918,-2068){\vector( 0, 1){233}}
}%
{\color[rgb]{0,0,0}\put(2035,-2068){\vector( 0, 1){233}}
}%
{\color[rgb]{0,0,0}\put(2152,-2068){\vector( 0, 1){233}}
}%
{\color[rgb]{0,0,0}\put(2269,-2068){\vector( 0, 1){233}}
}%
{\color[rgb]{0,0,0}\put(983,-2068){\vector( 0,-1){468}}
}%
{\color[rgb]{0,0,0}\multiput(633,-2068)(113.09677,0.00000){16}{\line( 1, 0){ 56.548}}
}%
{\color[rgb]{0,0,0}\put(4977,-961){\vector( 0, 1){233}}
}%
{\color[rgb]{0,0,0}\put(5094,-961){\vector( 0, 1){233}}
}%
{\color[rgb]{0,0,0}\put(5211,-961){\vector( 0, 1){233}}
}%
{\color[rgb]{0,0,0}\put(5328,-961){\vector( 0, 1){233}}
}%
{\color[rgb]{0,0,0}\put(5561,-961){\vector( 0, 1){233}}
}%
{\color[rgb]{0,0,0}\put(5678,-961){\vector( 0, 1){233}}
}%
{\color[rgb]{0,0,0}\put(5795,-961){\vector( 0, 1){233}}
}%
{\color[rgb]{0,0,0}\put(5912,-961){\vector( 0, 1){233}}
}%
{\color[rgb]{0,0,0}\put(4626,-961){\vector( 0, 1){467}}
}%
{\color[rgb]{0,0,0}\multiput(4276,-961)(113.09677,0.00000){16}{\line( 1, 0){ 56.548}}
}%
{\color[rgb]{0,0,0}\put(3042,-970){\vector( 0, 1){234}}
}%
{\color[rgb]{0,0,0}\put(3159,-970){\vector( 0, 1){234}}
}%
{\color[rgb]{0,0,0}\put(3276,-970){\vector( 0, 1){234}}
}%
{\color[rgb]{0,0,0}\put(3393,-970){\vector( 0, 1){234}}
}%
{\color[rgb]{0,0,0}\put(3626,-970){\vector( 0, 1){234}}
}%
{\color[rgb]{0,0,0}\put(3743,-970){\vector( 0, 1){234}}
}%
{\color[rgb]{0,0,0}\put(3860,-970){\vector( 0, 1){234}}
}%
{\color[rgb]{0,0,0}\put(3977,-970){\vector( 0, 1){234}}
}%
{\color[rgb]{0,0,0}\multiput(2341,-970)(113.09677,0.00000){16}{\line( 1, 0){ 56.548}}
}%
{\color[rgb]{0,0,0}\put(2458,-970){\vector( 0, 1){234}}
}%
{\color[rgb]{0,0,0}\put(2575,-970){\vector( 0, 1){234}}
}%
{\color[rgb]{0,0,0}\put(2808,-970){\vector( 0, 1){234}}
}%
{\color[rgb]{0,0,0}\put(2691,-970){\vector( 0,-1){234}}
}%
\put(3061,-241){\makebox(0,0)[lb]{\smash{\SetFigFont{12}{14.4}{sf}{\color[rgb]{0,0,0}(B)}%
}}}
\put(5041,-241){\makebox(0,0)[lb]{\smash{\SetFigFont{12}{14.4}{sf}{\color[rgb]{0,0,0}(C)}%
}}}
\put(901,-241){\makebox(0,0)[lb]{\smash{\SetFigFont{12}{14.4}{sf}{\color[rgb]{0,0,0}(A)}%
}}}
\put(1396,-2716){\makebox(0,0)[lb]{\smash{\SetFigFont{12}{14.4}{sf}{\color[rgb]{0,0,0}(D)}%
}}}
\put(4591,-2716){\makebox(0,0)[lb]{\smash{\SetFigFont{12}{14.4}{sf}{\color[rgb]{0,0,0}(E)}%
}}}
\put(4321,-1816){\makebox(0,0)[lb]{\smash{\SetFigFont{12}{14.4}{rm}{\color[rgb]{0,0,0}\amplone}%
}}}
\put(4276,-1366){\makebox(0,0)[lb]{\smash{\SetFigFont{12}{14.4}{rm}{\color[rgb]{0,0,0}\amplseven}%
}}}
\put(1126,-2401){\makebox(0,0)[lb]{\smash{\SetFigFont{12}{14.4}{rm}{\color[rgb]{0,0,0}\amplfive}%
}}}
\put(2791,-1231){\makebox(0,0)[lb]{\smash{\SetFigFont{12}{14.4}{rm}{\color[rgb]{0,0,0}\amplthree}%
}}}
\put(4681,-466){\makebox(0,0)[lb]{\smash{\SetFigFont{12}{14.4}{rm}{\color[rgb]{0,0,0}\amplfour}%
}}}
\put(5221,-601){\makebox(0,0)[lb]{\smash{\SetFigFont{12}{14.4}{rm}{\color[rgb]{0,0,0}\amplone}%
}}}
\put(2926,-601){\makebox(0,0)[lb]{\smash{\SetFigFont{12}{14.4}{rm}{\color[rgb]{0,0,0}\amplone}%
}}}
\put(991,-601){\makebox(0,0)[lb]{\smash{\SetFigFont{12}{14.4}{rm}{\color[rgb]{0,0,0}\amplone}%
}}}
\put(1666,-1726){\makebox(0,0)[lb]{\smash{\SetFigFont{12}{14.4}{rm}{\color[rgb]{0,0,0}\amplone}%
}}}
\end{picture}

\parbox[t]{8cm}{
\begin{enumerate}
\item[(A)] Start with the uniform superposition of the twelve states.
\item[(B)] Invert the amplitude of the target state.
\item[(C)] Invert about the average in each of the three blocks.
\item[(D)] Invert the amplitude of the target state again.
\item[(E)] Invert about the global average.
\end{enumerate}
}
\end{center}
\caption{Partial quantum search in a database of twelve items}
\label{fig:intro}
\end{figure}
Consider a list with twelve items.  It is easily checked that to find
the target with certainty, we would need at least three (quantum)
queries. Suppose we just want to know whether the given element
appears in the first one-third, the second one-third or the last
one-third. Can we now manage with fewer queries? Consider the
algorithm shown in Figure~\ref{fig:intro}. At the end, all the
amplitude is concentrated on the target block. So, by performing a
measurement, we can determine the target block with probability one,
and, in fact, recover the target state itself with probability
$\frac{3}{4}$. Note that this algorithm makes just two queries, for
the transitions from (A) to (B) and (C) to (D).

\section{Background and notation}

Before we describe our algorithm, we review the framework for
quantum search. We assume that the reader is familiar with the basics
of quantum circuits, especially the quantum database search algorithm
of Grover~\cite{Grover} (see, for example, Nielsen and
Chuang~\cite[Chapter 6]{NC}).

\subsection{Database search}

The database is modeled as a function $f: [N] \rightarrow S$, where
$[N] \overset{\Delta}{=} \{0,1,2,\ldots, N-1\}$ is the set of
addresses and $S$ is the set of possible items to be stored in the
database. For our purposes, we can take $S$ to be $\{0,1\}$. When
thinking of bits we identify $[N]$ with $\{0,1\}^n$. We will assume
that there is a unique location $x$ such that $f(x)=1$. This assumption
is natural and widely used when modeling the search problem for
unsorted databases with unique keys. We will call this the target
location and denote it by $t$. In the quantum model, this database is
provided to us by means of an oracle unitary transformation $T_f$
which acts on an $(n+1)$-qubit space by sending the basis vector $| x
\rangle| b \rangle$ to $| x \rangle | b \oplus f(x) \rangle $.

In the version of the quantum search algorithm useful for us (see
Nielsen and Chuang~\cite[Chapter 6]{NC}), one starts with
$n$-qubits in the uniform superposition
\[
| \psi_0 \rangle= \frac{1}{\sqrt{N}} \sum_{x \in \{0,1\}^n} | x \rangle.
\]
Then, one repeatedly applies the following operation in order to amplify the
probability of the target state: 
\[
\mathsf{A} \overset{\Delta}{=} I_0 I_t.
\]
Here, the unitary transformation $I_t= I - 2| t \rangle \langle t |$
selectively inverts the amplitude of the target state, and $I_0 = 2|
\psi_0 \rangle \langle \psi_0 | - I$ selectively inverts the amplitude
of the component orthogonal to $|\psi_0 \rangle$. One can implement
$I_t$ using a quantum circuit that makes one query to the oracle.  The
state vector always stays in the two dimensional space spanned by $|
\psi_0 \rangle$ and $| t \rangle$. Each application of the
transformation $\mathsf{A}$ moves the state vector towards the target
state by an angle of about $\frac{2}{\sqrt{N}}$. Thus, after
$(\frac{\pi}{4}) \sqrt{N}$ iterations the state vector reaches (very
close to) the target state. One curious feature of this algorithm is
that further applications of the transformation move the state vector
away from $| t \rangle$ and $| \psi_0 \rangle$ so that very soon its
component along the $| \psi_0 \rangle$ becomes negative.  Thus,
one has to choose the number of iteration carefully in order
to optimize the probability of success.  {\em Interestingly, this 
drift away from the target state, which is usually considered a nuisance, 
is crucial for our general  partial search algorithm. In fact, we
made use of it in the example above.}

\paragraph{Quantum algorithms that make no error:}
Grover's database search algorithm~\cite{Grover} finds the location
where an element is stored using approximately $\left( \frac{\pi
}{4}\right) \sqrt{N}$ queries to the oracle. The original version of
the algorithm could return a wrong address with probability $O(1/N)$.
(Recall that we assume that the element is stored at a unique location
in the database.) However, it is possible to modify the algorithm so
that the correct answer is returned with certainty (for example, one
can modify the last iteration slightly so that the state vector does
not overshoot its target see also~\cite{Brassard,Hu,Long,Zalka2}).  In
the next section, we present an algorithm for partial search that
returns the correct answer with high probability. It is easy to see
that this algorithm can be modified to ensure that the correct answer
is returned with probability one.  It is easy to reduce the usual
database problem to the partial search problem considered in this
paper and conclude from Zalka's~\cite{Zalka} lower bound of
$\left(\frac{\pi }{4}\right) \sqrt{N}$ that the bound we obtain are
essentially optimal for partial search algorithms that return the
correct answer with certainty.  A similar lower bound applies to
partial search algorithm whose probability of error is very small say
$O(N^{-\frac{1}{4}})$.  For this we need to observe that a version of
Zalka's~\cite{Zalka} lower bound holds even when the quantum database
search algorithm errs with some very small probability. A referee of
an earlier version of this paper pointed out that this fact is not
explicitly stated in Zalka's paper; so, we provide a detailed
derivation in Appendix~\ref{app:zalka}. In Section~\ref{sec:lb}, we
show that the savings achieved by our algorithm are essentially
optimal for algorithms that err with very small probability.

\subsection{The partial search problem}

In the partial search problem, the set of address, $R$, is partitioned
into $K$ equal blocks of size $N/K$ each. We will then think of the
addresses in $R$ as pairs $(y,z)$ where $y \in [K] \overset{\Delta}{=}
\{1,2, \ldots, K\}$ and $z \in [N/K]$. For example, when $R=\{0,1\}^n$
and $K=2^k$, each such block could correspond to those addresses that
have the same first $k$ bits; then, $y$ gives the first $k$ bits of
the address and $z$ the remaining $n-k$ bits. We refer to the block
containing the target $t=y_tz_t$ as the \emph{target} block and the
other blocks as \emph{non-target} blocks. In our algorithm we will
apply the database search iteration to the different blocks in
parallel. This is implemented by means of an operator
$\mathsf{A}_{[N/K]}$ which acts as follows. First, it applies the the
operator $I_t$ defined above, to invert the amplitude of target
state. Then, it performs an inversion about the average in each block
in parallel. Formally, let $| \psi_{0,[N/K]} \rangle
\overset{\Delta}{=} \frac{1}{\sqrt{N/K}} \sum_{z \in [N/K]} | z
\rangle$ and $I_{0,[N/K]} = 2 | \psi_{0,[N/K]} \rangle
\langle\psi_{0,[N/K]}| - I_{[N/K]}$ (where $I_{[N/K]}$ is the identity
operator on the space of dimension $N/K$). Then,
\[ \mathsf{A}_{[N/K]} = (I_{[K]} \otimes I_{0,[N/K]}) I_t.\]
Note this operator acts on the space of dimension $N$.  With this
notation, we are ready to describe our algorithm and the lower bound
in detail.

\section{Partial search is easier \ldots}
In this section, we present an algorithm that uses fewer queries to
the database than the algorithm presented in the introduction.  The
goal is to use a suitable combination of operators $\mathsf{A}$ and
$\mathsf{A}_{[N/K]}$ so that, in the end, only the basis states in the
target block have positive amplitude. Clearly, after this a
measurement of the state vector with respect to the standard basis
will give us the information we want.

The algorithm has three steps, of which the first two involve
amplitude amplification. In the first step, we use the operator
$\mathsf{A}$ on the entire address space (of dimension $N$). However,
we stop short of finding the target exactly, making
$\theta\left(\sqrt{\frac{N}{K}}\right)$ fewer queries than we would
for a complete search.  In the second step, we use the operator
$\mathsf{A}_{[N/K]}$ which does amplification in parallel in each
block. The amplitudes of the states in the non-target blocks are not
affected in this step; the states in the target block, however, acquire
negative amplitudes. The number of iterations is chosen so that the
overall average amplitude is exactly half of the amplitude of each
state in non-target blocks.  In particular, if we perform an inversion
about the average for all non-target states, the amplitude of the
states in the non-target blocks will become zero.  This inversion
about the average for non-target states is implemented in the third
step; it requires just one query to the database.  It will turn out
that the savings in the first step are significantly more than the
number of queries needed to implement the second step, so that in the
end we are still left with a savings of
$\theta\left(\sqrt{\frac{N}{K}}\right)$ queries.
\begin{figure}[tbp] \label{fig:alg}
\framebox{\parbox[t]{5.5in}{
\begin{quote}
{\bf Step 1:} Prepare the address register in the state
\[ \ket{\psi_0} = \frac{1}{\sqrt{N}} \sum_{x \in [N]} \ket{x}.\]
Perform $\ell_1(\epsilon) = (\frac{\pi}{4})(1-\epsilon) \sqrt{N}$ iterations of the standard
amplification step. The resulting state (see Figure {fig:step1})
is
\[ \ket{\psi_1} \defeq \amp^{\ell_1} \ket{\psi_0}.\]

\smallskip

{\bf Step 2:} We perform $\ell_2(\epsilon)$ iterations of the
$\amp_{[N/K]}$, so that the 
average amplitude of all the non-target states is exactly half the 
amplitude of every state in every non-target block. 

\smallskip

{\bf Step 3:} There are two operations performed in this step. First, we move the target
state out. That is, we take an ancilla qubit $b$ (initially in state $| 0
\rangle$) and perform the operation $M$: 
\begin{eqnarray*}
\mbox{for the target state $t$: \ \ }| b \rangle| t \rangle &\mapsto& | b
\oplus 1 \rangle | t \rangle; \\
\mbox{for other basis states $\ket{x}$: \ \ }| b \rangle| x \rangle
&\mapsto& | b \rangle| x \rangle.
\end{eqnarray*}
Controlled on $b=0$, we perform an inversion about the average. All
states in the non-target blocks now have amplitude zero!
\end{quote}
}
} 
\caption{The algorithm for partial search}
\end{figure}

The algorithm in Figure~\ref{fig:alg} uses a parameter $\epsilon$ that
controls the number of iterations. Later, in order to minimize the
number of queries we will need to choose $\epsilon$ optimally.  On
first reading, it might be helpful to assume that $K$ is a large
constant and $\epsilon = \frac{1}{\sqrt{K}}$.

\begin{figure}[tbp]
\begin{center}
\input{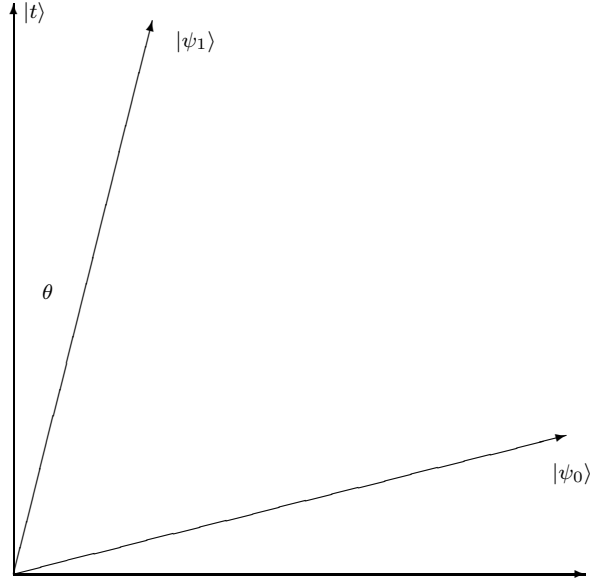}
\end{center}
\caption{Step 1 moves the state vector close to the target state by performing
$\frac{\pi}{4}(1-\epsilon)\sqrt{N}$ iterations of the standard quantum search}
\label{fig:step1}
\end{figure}

\subsection{Estimating the number of queries}

\paragraph{Remark about approximations:}
In the following we will assume that $N$ is large. To keep the
presentation simple, we will often make approximate calculations. When
we say $\mbox{LHS} \sim \mbox{RHS}$, we mean that the two sides differ
by a quantity that goes to $0$ as $N$ becomes large, say like
$O(\frac{1}{\sqrt{N}})$. For example, we sometimes pretend that the
target state is orthogonal to the state $| \psi_0 \rangle$. Our
algorithm will produce the correct answer with probability very close
to 1; the probability of error is
$O\left(\frac{1}{\sqrt{N}}\right)$. However, as remarked above, one
can modify the algorithm so that it returns the correct answer with
certainty.

\begin{figure}[t]
\begin{center}
\setlength{\unitlength}{4144sp}%
\begingroup\makeatletter\ifx\SetFigFont\undefined
\def\x#1#2#3#4#5#6#7\relax{\def\x{#1#2#3#4#5#6}}%
\expandafter\x\fmtname xxxxxx\relax \def\y{splain}%
\ifx\x\y   
\gdef\SetFigFont#1#2#3{%
  \ifnum #1<17\tiny\else \ifnum #1<20\small\else
  \ifnum #1<24\normalsize\else \ifnum #1<29\large\else
  \ifnum #1<34\Large\else \ifnum #1<41\LARGE\else
     \huge\fi\fi\fi\fi\fi\fi
  \csname #3\endcsname}%
\else
\gdef\SetFigFont#1#2#3{\begingroup
  \count@#1\relax \ifnum 25<\count@\count@25\fi
  \def\x{\endgroup\@setsize\SetFigFont{#2pt}}%
  \expandafter\x
    \csname \romannumeral\the\count@ pt\expandafter\endcsname
    \csname @\romannumeral\the\count@ pt\endcsname
  \csname #3\endcsname}%
\fi
\fi\endgroup
\begin{picture}(6466,4240)(5266,-3504)
\thinlines
{\color[rgb]{0,0,0}\put(5698,-724){\vector( 0, 1){356}}
}%
{\color[rgb]{0,0,0}\put(6307,-724){\vector( 0, 1){356}}
}%
{\color[rgb]{0,0,0}\put(6180,-724){\vector( 0, 1){356}}
}%
{\color[rgb]{0,0,0}\put(5926,-724){\vector( 0, 1){356}}
}%
{\color[rgb]{0,0,0}\put(5800,-724){\vector( 0, 1){356}}
}%
{\color[rgb]{0,0,0}\put(6409,-724){\vector( 0, 1){356}}
}%
{\color[rgb]{0,0,0}\put(6866,-724){\vector( 0, 1){356}}
}%
{\color[rgb]{0,0,0}\put(7476,-724){\vector( 0, 1){356}}
}%
{\color[rgb]{0,0,0}\put(7349,-724){\vector( 0, 1){356}}
}%
{\color[rgb]{0,0,0}\put(7095,-724){\vector( 0, 1){356}}
}%
{\color[rgb]{0,0,0}\put(6968,-724){\vector( 0, 1){356}}
}%
{\color[rgb]{0,0,0}\put(7577,-724){\vector( 0, 1){356}}
}%
{\color[rgb]{0,0,0}\put(7222,-724){\vector( 0, 1){356}}
}%
{\color[rgb]{0,0,0}\put(7933,-724){\vector( 0, 1){356}}
}%
{\color[rgb]{0,0,0}\put(8542,-724){\vector( 0, 1){356}}
}%
{\color[rgb]{0,0,0}\put(8416,-724){\vector( 0, 1){356}}
}%
{\color[rgb]{0,0,0}\put(8161,-724){\vector( 0, 1){356}}
}%
{\color[rgb]{0,0,0}\put(8035,-724){\vector( 0, 1){356}}
}%
{\color[rgb]{0,0,0}\put(8644,-724){\vector( 0, 1){356}}
}%
{\color[rgb]{0,0,0}\put(8288,-724){\vector( 0, 1){356}}
}%
{\color[rgb]{0,0,0}\put(8974,-724){\vector( 0, 1){356}}
}%
{\color[rgb]{0,0,0}\put(9584,-724){\vector( 0, 1){356}}
}%
{\color[rgb]{0,0,0}\put(9457,-724){\vector( 0, 1){356}}
}%
{\color[rgb]{0,0,0}\put(9203,-724){\vector( 0, 1){356}}
}%
{\color[rgb]{0,0,0}\put(9076,-724){\vector( 0, 1){356}}
}%
{\color[rgb]{0,0,0}\put(9685,-724){\vector( 0, 1){356}}
}%
{\color[rgb]{0,0,0}\put(9330,-724){\vector( 0, 1){356}}
}%
{\color[rgb]{0,0,0}\put(10727,-724){\vector( 0, 1){356}}
}%
{\color[rgb]{0,0,0}\put(11336,-724){\vector( 0, 1){356}}
}%
{\color[rgb]{0,0,0}\put(11209,-724){\vector( 0, 1){356}}
}%
{\color[rgb]{0,0,0}\put(10955,-724){\vector( 0, 1){356}}
}%
{\color[rgb]{0,0,0}\put(10828,-724){\vector( 0, 1){356}}
}%
{\color[rgb]{0,0,0}\put(11438,-724){\vector( 0, 1){356}}
}%
{\color[rgb]{0,0,0}\put(11082,-724){\vector( 0, 1){356}}
}%
\thicklines
{\color[rgb]{0,0,0}\put(5672,-724){\line( 1, 0){5842}}
}%
\thinlines
{\color[rgb]{0,0,0}\put(6053,-724){\vector( 0, 1){1448}}
}%
\put(5698,-927){\makebox(0,0)[lb]{\smash{\SetFigFont{7}{8.4}{rm}{\color[rgb]{0,0,0}TARGET BLOCK}%
}}}
\put(8110,-927){\makebox(0,0)[lb]{\smash{\SetFigFont{7}{8.4}{rm}{\color[rgb]{0,0,0}NON-TARGET  BLOCKS}%
}}}
{\color[rgb]{0,0,0}\put(6866,-2476){\vector( 0, 1){356}}
}%
{\color[rgb]{0,0,0}\put(7476,-2476){\vector( 0, 1){356}}
}%
{\color[rgb]{0,0,0}\put(7349,-2476){\vector( 0, 1){356}}
}%
{\color[rgb]{0,0,0}\put(7095,-2476){\vector( 0, 1){356}}
}%
{\color[rgb]{0,0,0}\put(6968,-2476){\vector( 0, 1){356}}
}%
{\color[rgb]{0,0,0}\put(7577,-2476){\vector( 0, 1){356}}
}%
{\color[rgb]{0,0,0}\put(7222,-2476){\vector( 0, 1){356}}
}%
{\color[rgb]{0,0,0}\put(7933,-2476){\vector( 0, 1){356}}
}%
{\color[rgb]{0,0,0}\put(8542,-2476){\vector( 0, 1){356}}
}%
{\color[rgb]{0,0,0}\put(8416,-2476){\vector( 0, 1){356}}
}%
{\color[rgb]{0,0,0}\put(8161,-2476){\vector( 0, 1){356}}
}%
{\color[rgb]{0,0,0}\put(8035,-2476){\vector( 0, 1){356}}
}%
{\color[rgb]{0,0,0}\put(8644,-2476){\vector( 0, 1){356}}
}%
{\color[rgb]{0,0,0}\put(8288,-2476){\vector( 0, 1){356}}
}%
{\color[rgb]{0,0,0}\put(8974,-2476){\vector( 0, 1){356}}
}%
{\color[rgb]{0,0,0}\put(9584,-2476){\vector( 0, 1){356}}
}%
{\color[rgb]{0,0,0}\put(9457,-2476){\vector( 0, 1){356}}
}%
{\color[rgb]{0,0,0}\put(9203,-2476){\vector( 0, 1){356}}
}%
{\color[rgb]{0,0,0}\put(9076,-2476){\vector( 0, 1){356}}
}%
{\color[rgb]{0,0,0}\put(9685,-2476){\vector( 0, 1){356}}
}%
{\color[rgb]{0,0,0}\put(9330,-2476){\vector( 0, 1){356}}
}%
{\color[rgb]{0,0,0}\put(10727,-2476){\vector( 0, 1){356}}
}%
{\color[rgb]{0,0,0}\put(11336,-2476){\vector( 0, 1){356}}
}%
{\color[rgb]{0,0,0}\put(11209,-2476){\vector( 0, 1){356}}
}%
{\color[rgb]{0,0,0}\put(10955,-2476){\vector( 0, 1){356}}
}%
{\color[rgb]{0,0,0}\put(10828,-2476){\vector( 0, 1){356}}
}%
{\color[rgb]{0,0,0}\put(11438,-2476){\vector( 0, 1){356}}
}%
{\color[rgb]{0,0,0}\put(11082,-2476){\vector( 0, 1){356}}
}%
{\color[rgb]{0,0,0}\multiput(6557,-2323)(13.51571,0.00000){383}{\makebox(1.5875,11.1125){\SetFigFont{5}{6}{rm}.}}
}%
\thicklines
{\color[rgb]{0,0,0}\put(5672,-2476){\line( 1, 0){5842}}
}%
\thinlines
{\color[rgb]{0,0,0}\put(5698,-2476){\vector( 0,-1){1016}}
}%
{\color[rgb]{0,0,0}\put(6409,-2476){\vector( 0,-1){1016}}
}%
{\color[rgb]{0,0,0}\put(6307,-2476){\vector( 0,-1){1016}}
}%
{\color[rgb]{0,0,0}\put(6206,-2476){\vector( 0,-1){1016}}
}%
{\color[rgb]{0,0,0}\put(5952,-2476){\vector( 0,-1){1016}}
}%
{\color[rgb]{0,0,0}\put(5825,-2476){\vector( 0,-1){1016}}
}%
{\color[rgb]{0,0,0}\put(6079,-2501){\vector( 0, 1){888}}
}%
\put(8110,-2679){\makebox(0,0)[lb]{\smash{\SetFigFont{7}{8.4}{rm}{\color[rgb]{0,0,0}NON-TARGET  BLOCKS}%
}}}
\put(5266,-2400){\makebox(0,0)[lb]{\smash{\SetFigFont{7}{8.4}{rm}{\color[rgb]{0,0,0}TARGET BLOCK}%
}}}
\end{picture}
\end{center}
\caption{After Step 2, the non-target states in the target block acquire
negative amplitudes}
\label{fig:histogram}
\end{figure}
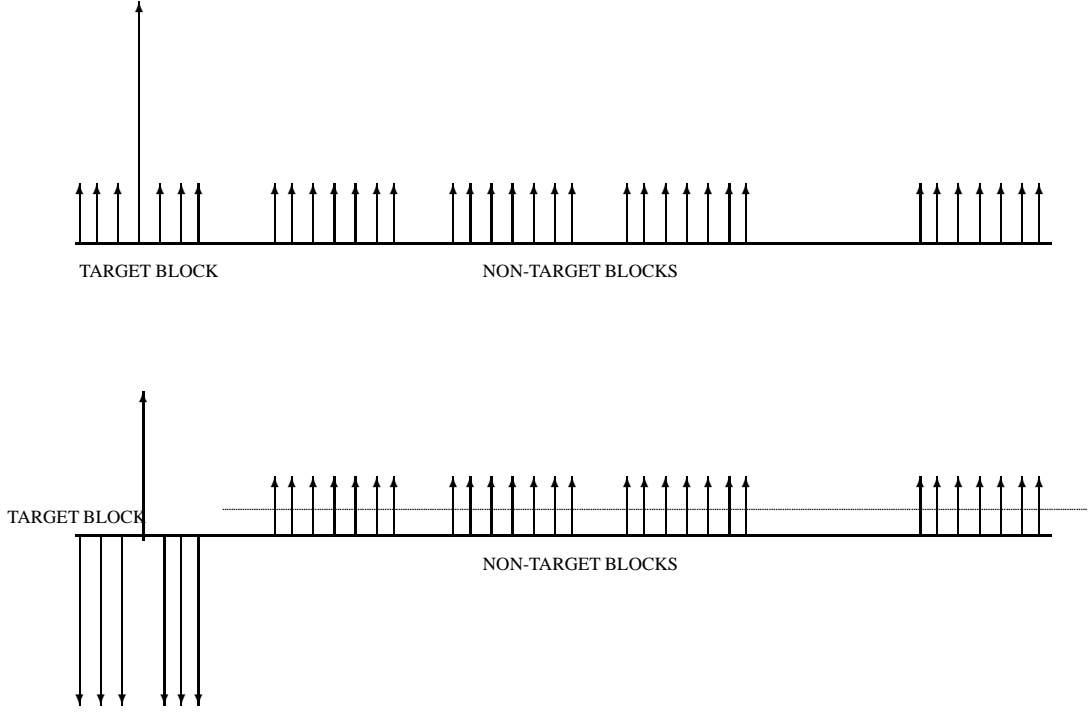

\paragraph{Analysis of Step 1:}
We will write $x=yz$ with $y \in [K]$ and $z \in [N/K]$, as described above,
and express $| \psi_1 \rangle$ as 
\[
\sum_{y \in [K]} \alpha_y | y \rangle | \psi_y \rangle,
\]
where the $\alpha_y$'s are complex amplitudes whose squares sum to $1$, and $%
| \psi_y \rangle$ is a state of the form 
\[
\sum_{z \in [N/K]} \beta_{yz} | z \rangle \mbox{ where } \sum_{z \in [N/K]}
|\beta_{yz}|^2 =1.
\]
Since, 
\[
| \psi_1 \rangle = \cos(\theta) | t \rangle + \frac{\sin (\theta)}{\sqrt{N%
}}\sum_{x: x \neq t} | x \rangle,
\]
we get the following information about $\alpha_y$ and $| \psi_y
\rangle$.  (See also the first histogram of
Figure~\ref{fig:histogram}). Recall that we write $t=y_tz_t$, where
$y_t$ is the address of the target block and $z_t$ is the address
inside the target block.

\begin{description}
\item[In a non-target block:] Suppose $y \neq y_t$. Then, the
projection of the current state vector $| \psi_1 \rangle$ on the
subspace corresponding the block of $y$, is a uniform superposition of
the basis vectors but with length $\frac{1}{\sqrt{K}}
\sin(\theta)$. That is,
\begin{eqnarray}
\alpha_y &\sim & \frac{1}{\sqrt{K}} \sin(\theta);  \nonumber \\
\mbox{ and } | \psi_y \rangle &=& \frac{1}{\sqrt{N/K}} \sum_{z \in [N/K]} |
z \rangle.  \label{eq:non-target}
\end{eqnarray}

\item[In the target block:] 
For the target block, we have 
\begin{eqnarray}
\alpha_{y_t} &\sim & \sqrt{1-\left(\frac{K-1}{K}\right) \sin^2(\theta)}\ ; 
\nonumber \\
\mbox{ and\ \  } | \psi_{y_t} \rangle &\sim & \frac{\cos(\theta)}{\alpha_{y_t}
}|z_t\rangle + \frac{\sin(\theta)}{\alpha_{y_t} \sqrt{K}}
\frac{1}{\sqrt{N/K}}
 \sum_{z \in [N/K]} | z \rangle.  \label{eq:target}
\end{eqnarray}
The first term corresponds to the component along the target state; the
second corresponds to the uniform superposition on all states in the target
block.
\end{description}

\paragraph{Analysis of Step 2:}

In this step, we work on each block separately but in parallel by applying
the operator $A_{[N/K]}$ described above. This operator has the following
effect. The amplitudes of basis states in non-target blocks remain the same
(i.e.~$\frac{\sin(\theta)}{\sqrt{N}}$). However, the non-target basis states
in the target block transfer their amplitude to the target state and then
acquire negative amplitudes. For $y \neq y_t$, 
$\alpha_y$ and $|\psi_y\rangle$ do not change. However, the projection
on the subspace corresponding to the target block moves from
$|\psi_{y_t}\rangle$ to $\psi'_{y_t}$ (see
Figure~\ref{fig:step2}).  Thus, at the end of Step 2, the
overall state is
\[
| \psi_2 \rangle = \alpha_{y_t} | y_t \rangle | \psi_{y_t}' \rangle + 
\sum_{y: y \neq y_t} \alpha_y | y \rangle  | \psi_y \rangle.
\]
We choose $\ell_2$ so that the component of the vector $| \psi_{y_t}'
\rangle$ along $\frac{1}{\sqrt{N/K}}\sum_{z \in [N/K]} | z \rangle$
is an appropriate negative quantity.
The histogram in Figure~\ref{fig:histogram} shows the amplitudes of all the
basis states before and after Step 2. The dotted line in the histogram is
the average of all the non-target states and is arranged to be half the
amplitude of the states in the non-target blocks. 

\begin{figure}[tbp]
\begin{center}
\setlength{\unitlength}{4144sp}%
\begingroup\makeatletter\ifx\SetFigFont\undefined
\def\x#1#2#3#4#5#6#7\relax{\def\x{#1#2#3#4#5#6}}%
\expandafter\x\fmtname xxxxxx\relax \def\y{splain}%
\ifx\x\y   
\gdef\SetFigFont#1#2#3{%
  \ifnum #1<17\tiny\else \ifnum #1<20\small\else
  \ifnum #1<24\normalsize\else \ifnum #1<29\large\else
  \ifnum #1<34\Large\else \ifnum #1<41\LARGE\else
     \huge\fi\fi\fi\fi\fi\fi
  \csname #3\endcsname}%
\else
\gdef\SetFigFont#1#2#3{\begingroup
  \count@#1\relax \ifnum 25<\count@\count@25\fi
  \def\x{\endgroup\@setsize\SetFigFont{#2pt}}%
  \expandafter\x
    \csname \romannumeral\the\count@ pt\expandafter\endcsname
    \csname @\romannumeral\the\count@ pt\endcsname
  \csname #3\endcsname}%
\fi
\fi\endgroup
\begin{picture}(2072,2702)(7312,-3848)
\thinlines
{\color[rgb]{0,0,0}\put(8745,-3836){\vector( 0, 1){2678}}
}%
{\color[rgb]{0,0,0}\put(8745,-3836){\vector(-2, 3){1485.846}}
}%
{\color[rgb]{0,0,0}\put(8742,-3836){\vector( 1, 4){648.823}}
}%
\put(8789,-1226){\makebox(0,0)[lb]{\smash{\SetFigFont{8}{9.6}{rm}{\color[rgb]{0,0,0}$\ket{z_t}$}%
}}}
\put(8786,-2810){\makebox(0,0)[lb]{\smash{\SetFigFont{8}{9.6}{rm}{\color[rgb]{0,0,0}$\theta_1$}%
}}}
\put(8406,-3125){\makebox(0,0)[lb]{\smash{\SetFigFont{8}{9.6}{rm}{\color[rgb]{0,0,0}$\theta_2$}%
}}}
\put(7397,-1611){\makebox(0,0)[lb]{\smash{\SetFigFont{8}{9.6}{rm}{\color[rgb]{0,0,0}$\ket{\psi_{y_t}'}$}%
}}}
\put(9384,-1400){\makebox(0,0)[lb]{\smash{\SetFigFont{8}{9.6}{rm}{\color[rgb]{0,0,0}$\ket{\psi_{y_t}}$}%
}}}
\end{picture}
\end{center}
\caption{Step 2 consists of independent quantum searches
for each block; in the target block the state vector moves past the target}
\label{fig:step2}
\end{figure}
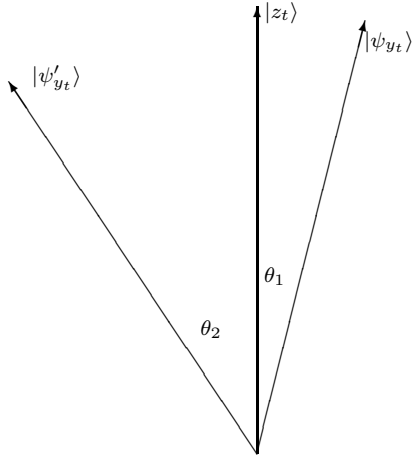


%

We are now ready to determine $\ell_2$. We work in the $N/K$
dimensional space corresponding to the target block. Let $\theta_1$ be
the initial angle between $| \psi_{y_t} \rangle$ and $|z_t
\rangle$. Then, by (\ref{eq:target}), $\sin(\theta_1) = \frac{1}{%
\alpha_{y_t}\sqrt{K}} \sin (\theta)$, that is,
\begin{equation}
\theta_1 = \arcsin\left(\frac{1}{\alpha_{y_t}\sqrt{ K}} \sin \theta\right).
\label{eq:theta1}
\end{equation}
Let $\theta_2$ be the angle between the final state $| \psi'_{y_t} \rangle$
and $|z_t \rangle$ that we want to achieve in Step~2. To determine $\theta_2
$ we need to do a small calculation. Let $X$ be the sum of the amplitudes of
all the states in non-target blocks. Then, we have from (\ref{eq:non-target}%
) that 
\[
X \sim \frac{(K-1)\sqrt{N}}{K} \sin \theta.
\]
Let $Y$ denote the sum of the amplitudes of all non-target states in the
target block at the end of Step~2. We want the overall average amplitude
per block (that is, $(X+Y)/K$) to be half the average in the non-target blocks
(that is, $X/(2(K-1))$); therefore,
\[
Y \sim -\frac{K-2}{2(K-1)} X \sim - \frac{(K-2)\sqrt{N}}{2K}\sin \theta.
\]
On the other hand, $Y= -\alpha_{y_t}\sin(\theta_2)\sqrt{N/K}$. Thus, 
\begin{equation}
\theta_2 \sim \arcsin\left (\frac{1}{2\alpha_{y_t}} \frac{K-2}{\sqrt{K}}
\sin \theta\right).  \label{eq:theta2}
\end{equation}

The vector $| \psi_{y_t} \rangle$ needs to traverse a total angle of $%
\theta_1+\theta_2$. In one iteration of $A_{[N/K]}$ it covers $\frac{2}{%
\sqrt{N/K}}$. We thus have an estimate for $\ell_2$: 
\[
\ell_2(\epsilon) = \frac{\sqrt{N/K}}{2}(\theta_1+\theta_2),
\]
where (\ref{eq:theta1}) and (\ref{eq:theta2}) give us the values of $\theta_1
$ and $\theta_2$ in terms of $\epsilon$. We choose $\epsilon$ so that the
total number of queries, namely, 
\begin{eqnarray*}
(\ell_1(\epsilon) + \ell_2(\epsilon)) \sqrt{N} +1 &\sim & \left(\frac{\pi}{4}%
\right)(1-\epsilon) \sqrt{N}+ \frac{1}{2 \sqrt{K}} \left[
 \arcsin\left(\frac{1}{%
\alpha_{y_t}\sqrt{ K}} \sin \theta\right) \right. \\
&& \mbox{\ \ \hspace{1in} \ \ \ \ \ \ \ } {} + \left. \arcsin\left (\frac{1}{%
2\alpha_{y_t}} \frac{K-2}{\sqrt{K}} \sin \theta\right)\right] \sqrt{N} + 1
\end{eqnarray*}
is as small as possible.

Note that with $\epsilon =0$, the algorithm reduces to the usual database
search algorithm and takes $(\frac{\pi}{4}) \sqrt{N})$ queries. Also, for $%
K\geq 2$, the derivative (w.r.t. $\epsilon$) of 
\[
\ell_1(\epsilon) + \ell_2(\epsilon) ,
\]
is negative at $\epsilon=0$. It follows that with the optimum choice of $%
\epsilon$, the running time is bounded by $\frac{\pi}{4}(1 - c_k) \sqrt{N}$
for some constant $c_k > 0$.

We have not been able to obtain general expression for the optimum number of
queries by an optimum choice of $\epsilon$. For some small values of $K$,
the following table lists the optimum values obtained by using a computer
program.

\begin{center}
\begin{tabular}{||l|l|l||}
\hline
& Upper bound & Lower bound \\ \hline
Database search & $0.785 \sqrt{N} $ & $0.785 \sqrt{N} $ \\ \hline
K=2 (first bit) & $0.555 \sqrt{N} $ & $0.23 \sqrt{N} $ \\ \hline
K=3 (first trit) & $0.592 \sqrt{N} $ & $0.332 \sqrt{N} $ \\ \hline
K=4 (first two bits) & $0.615 \sqrt{N} $ & $0.393 \sqrt{N} $ \\ \hline
K=5 & $0.633 \sqrt{N} $ & $0.434 \sqrt{N} $ \\ \hline
K=8 (first three bits) & $0.664 \sqrt{N} $ & $0.508 \sqrt{N}$ \\ \hline
K=32 (first five bits) & $0.725 \sqrt{N} $ & $0.647 \sqrt{N} $ \\ \hline
\end{tabular}
\end{center}

For large values $K$, one can obtain some estimates on the behavior of
$c_k$. For example, assume that $K$ is large (and $N$ even
larger). Take $\epsilon = \frac{1}{\sqrt{K}}$. Then, we approximate
$\sin \theta$ by $\theta= \frac{\pi}{2 \sqrt{K}}$ and bound the number
of queries by
\[
\frac{\pi}{4}\left[ 1 - \left(1 - \frac{2}{\pi} \arcsin \frac{\pi}{4}\right)
\frac{1}{\sqrt{K}} + O \left(\frac{1}{K}\right)\right]\sqrt{N} +1 \leq \frac{
\pi}{4}\left( 1 - \frac{0.42}{\sqrt{K}} \right) \sqrt{N} . 
\]

\section{\ldots but not much easier}

\label{sec:lb}

\begin{theorem} Fix $K$. Suppose for all large $N$, there is an algorithm
for the partial search problem that uses at most $\alpha_K \sqrt{N}$
queries and return the correct answer with probability of error
$O(N^{-\frac{1}{4}})$. Then,
\[ \alpha_K \geq \left(\frac{\pi}{4}\right)\left(1-\frac{1}{\sqrt{K}}\right).\]
\end{theorem}
\begin{proof} 
We reduce the database search problem to the partial search problem.
First, we derive the lower bound assuming that the algorithms for the
partial search problem return the correct answer with certainty.
We start by applying the algorithm for partial search for databases of
size $N$. This yields the first $\log K$ bits of the target state.
Next, we restrict ourselves to those addresses $x$ that have the
correct first $k$ bits and determine the next $k$ bits of the universe
by using the partial search algorithm for databases of size
$N/K$. Continuing in this way, we converge on the target state after
making a total of at most
\[ \alpha \left(1+\frac{1}{\sqrt{K}} + \frac{1}{K}
+ \frac{1}{K\sqrt{K}} + \cdots  \right) \leq \alpha_k
\left(\frac{\sqrt{K}}{\sqrt{K} - 1}\right)
\sqrt{N}
\]
queries to the database. Thus, using Zalka's lower bound~\cite{Zalka}, we have
\[ \alpha_k
\left(\frac{\sqrt{K}}{\sqrt{K} - 1}\right)
\sqrt{N}
\geq \left(\frac{\pi}{4}\right) \sqrt{N},\]
implying that 
\[ \alpha_k \geq
\left(\frac{\pi}{4}\right)\left(1-\frac{1}{\sqrt{K}}\right).\]

The reduction for algorithms with errors is similar. We apply the
above method until the problem size becomes very small (say less than
$N^{\frac{1}{3}}$).  At that point we solve the database search
problem by brute force.  Since, we invoked the partial search problem
only $O(\log N)$ times, and each such invocation had probability of
error at most $N^{-\frac{1}{12}}$, the probability that all these
invocations gave the correct answer is $1-O( N^{-\frac{1}{12}}\log
N)$. This gives a database search algorithm that gives the correct
answer with probability $1-O( N^{-\frac{1}{12}}\log N)$.  Now,
Theorem~\ref{thm:zalka-error} in the appendix implies that such an
algorithm must make $\frac{\pi}{4}\sqrt{N} (1-o(1))$ queries to the
oracle. A calculation very similar to the one above gives the claimed
lower bound for $\alpha_K$.  \qed
\end{proof}

\section{Summary}

In yet another variant of the quantum search problem we ask whether it
is any easier to partially search a database than it is to completely
search it. This question is significant, because the search algorithm
of \cite{Grover} has been shown in \cite{Zalka} to be precisely
optimal for the complete search problem. The short answer is that
partial search is slightly easier. We derive a lower bound indicating
how much easier it is and an algorithm that closely matches this lower
bound. In the appendix, we present a lower bound on the expected
number of queries made by zero-error randomized algorithms for partial
search; we also derive an explicit form of Zalka's lower bound for
quantum search algorithms that make very small error.

\subsection*{Acknowledgments}

We thank the referees of on an earlier version of this paper for their
detailed comments.

\appendix

\section{Randomized complexity of partial search}
\label{app:randomized}
We are considering algorithms that make no error.  To show a lower
bound of $t$ queries on the expected number of probes made by a
randomized algorithm on a worst-case input, it is enough to describe a
distribution on inputs for which every deterministic algorithm makes
at least $t$ queries on the average.

 Let $\ell_1, \ell_2, \ldots,$ be the sequence of location this
deterministic algorithm probes if $f(x)=0$, for all $x$. Now, consider
the uniform distribution on inputs obtained by choosing a random
location $t \in [N]$ and setting $f(x)=1$ if and only $x=t$. Let
${\cal E}$ denote the event that $t$ is one of $\ell_1,\ell_2,\ldots,
\ell_{N-N/K}$. Then, the probability of ${\cal E}$ is exactly $1 -
1/K$ and conditioned on the event ${\cal E}$, the expected number of
probes made by the deterministic algorithms is exactly
$\frac{N}{2}\left(1 -\frac{1}{K}\right)$. If ${\cal E}$ does not hold
then, the algorithm makes at least $N\left(1 -\frac{1}{K}\right)$
probes (because it is not allowed any errors). Thus, the overall
average number of probes made by this algorithm is at least
\[ \left(1 - \frac{1}{K}\right) \times \frac{N}{2}\left(1 -\frac{1}{K}\right)
+ \frac{1}{K}\times N\left(1 -\frac{1}{K}\right) = \frac{N}{2}\left(1
-\frac{1}{K^2}\right).\]

\section{Zalka's bound revisited}

\label{app:zalka}

In this section we give a detailed proof of the following theorem.

\begin{theorem} \label{thm:zalka-error}
Suppose a quantum database search algorithm makes $T$ queries to a
database of size $N\geq 100$ and returns the correct answer with
probability at least $1-\epsilon \geq 0.9$.  Then,
\[ T \geq \frac{\pi}{4} \left( 1 - O\left(\sqrt{\epsilon} + 
N^{-\frac{1}{4}}\right)\right).\] 
\end{theorem}
To justify this, we will make the arguments in Zalka's
paper~\cite{Zalka} more explicit and rigorous. Let $O_y$ be the oracle
corresponding to the database where $y\in [N]$ is the only marked
element.  For $t=0,1,\ldots,T$, let
\begin{description}
\item[$\ket{\phi_t}$] be the state of the registers just before the
$(t+1)$-st query is made assuming that all instances of the oracle 
are replaced by the identity transformation;
\item[$\ket{\phi^y_t}$] be the state of the registers just before the
$(t+1)$-st query is made to the oracle, assuming that the oracle is $O_y$.
\end{description}
Thus, $\ket{\phi_0}=\ket{\phi^y_0}$ is the initial state and
$\ket{\phi_T}$ and $\ket{\phi^y_T}$ are the states reached at the
end. Like several other lower bound proofs, our proof will also
consider the variation in the state caused by each query to the
oracle. To state this, we will denote by $\ket{\phi^{y,i}_t}$ the
state of the registers just before the $(t+1)$-st query is made
assuming that the first $T-i$ queries are made to the identity oracle
and the remaining $i$ queries are made to the oracle $O_y$. Thus,
$\ket{\phi^{y,0}_t}=\ket{\phi_t}$ and $\ket{\phi^{y,T}_t} =
\ket{\phi_t^y}$.

It will be convenient to have only real valued amplitudes appearing in
our state vectors when they are expressed in the computational basis.
The standard way to arrange this is by adding an extra qubit register
and associating the component $(a+ib)\ket{x}$ appearing in the
original space with the vector $a\ket{x}\ket{0} + b
\ket{x}\ket{1}$. We will assume that the first register (whose state
is a superposition of values in $[N]$, when we identify $[N]$ with
$\{0,1\}^n$, this register will consist of $n$ qubits) contains the
query made to the oracle and also that this {\em address} register is
measured at the end to recover the answer. Let $V$ be the Hilbert
space associated with the remaining registers, and let $P_y$ be the
orthogonal projection operator for the subspace $\ket{y} \tensor V$.
Note that $\|P_y \ket{\phi}\|^2$ is the probability that we observe
$y$ when we measure the first register in the state $\ket{\phi}$.

As in Zalka's argument, we will use angles to measure the distance
between (real) unit vectors.  Let $\theta(\phi, \phi')= \arccos
|\braket{\phi}{\phi'}|$. Note that this distance is always in the
range $[0,\frac{\pi}{2}]$, and that the triangle inequality holds for
it.  Our proof rests on three lemmas.

\begin{lemma}\label{lm:largeaveragedistance}
$\displaystyle
\sum_{y \in [N]} \theta(\phi_T,\phi^y_T) \geq \frac{\pi}{2}\left(1 -
O\left(\sqrt{\epsilon}+ N^{-\frac{1}{4}}\right)\right).$
\end{lemma}

\begin{lemma}\label{lm:smallpairwisedistance}
For $i=1,2,\ldots,T$,
$\theta(\phi^{y,i-1}_T, \phi^{y,i}_T) \leq 2\arcsin \sqrt{p_{T-i,y}}$,
where $p_{T-i,y} =\|P_y \ket{\phi_{T-i}}\|^2$ is the probability that the
first address register of $\ket{\phi_{T-i}}$ contain the value $y$.
\end{lemma}

Let us assume that these lemmas hold and proceed with the proof of our
theorem. Then,
\begin{eqnarray*}
\sum_{y \in [N]}\sum_{i=0}^{T-1} 2\arcsin \sqrt{p_{i,y}} 
&\geq &
\sum_{y \in [N]}\sum_{i=0}^{T-1} \theta(\phi_T^{y,i},\phi_T^{y,i+1}) 
\mbox{\ \ \ \ \ \  (by Lemma~\ref{lm:smallpairwisedistance})}
\\
&\geq & \sum_{y \in [N]} \theta(\phi_T^{y,0}, \phi_T^{y,T})
\mbox{\ \ \ \ \ \ (triangle inequality)}
\\
&\geq & \sum_{y \in [N]} \theta(\phi_T, \phi_T^y)\\
&\geq &N\cdot \frac{\pi}{2}\left(1 -
O\left(\sqrt{\epsilon}+ N^{-\frac{1}{4}}\right)\right).
\mbox{\ \ \ \ \ \ (by Lemma~\ref{lm:largeaveragedistance})}
\end{eqnarray*}
On changing the order of summation, we obtain
\begin{equation} \sum_{i=0}^{T-1} \sum_{y \in [N]} 2\arcsin
  \sqrt{p_{i,y}} \geq 
N\cdot \frac{\pi}{2}\left(1 -
O\left(\sqrt{\epsilon}+ N^{-\frac{1}{4}}\right)\right).
 \label{eq:doublesum}
\end{equation} 
Let us consider the inner sum for each value of $i$. 
\begin{lemma} \label{lm:equalprobability}
For $i=0,1,\ldots,T-1$,
$\displaystyle \sum_{y \in [N]} \arcsin \sqrt{p_{i,y}} = 
\sqrt{N} \left(1+O\left(\frac{1}{N}\right)\right)$. 
\end{lemma}
Now, our theorem follows immediately by using this bound in
(\ref{eq:doublesum}). It remains to prove the lemmas.

\subsection{Proofs of the lemmas}

\paragraph{Proof of Lemma~\ref{lm:largeaveragedistance}:}
Recall that $\theta(\phi_T, \phi_T^y) = \arccos |
\braket{\phi_T}{\phi_T^y} |$, and $P_y$ is the orthogonal projection
operator onto the subspace $\ket{y} \tensor V$. Since
$\ket{\phi_T}$ is a unit vector, we have
$\sum_y \| P_y \ket{\phi_T} \|^2 \leq 1$,
implying
\[ \frac{1}{N} \sum \|P_y \ket{\phi_T}\| \leq \left(\frac{1}{N} 
\sum_y P_y \ket{\phi_T} \|^2\right)^{1/2} \leq \frac{1}{\sqrt{N}}.\]
By Markov's inequality, for all but a fraction $N^{-\frac{1}{4}}$ of
the $y$'s in $[N]$,
\begin{equation}
 \|P_y \ket{\phi_T}\| \leq N^{-\frac{1}{4}}.
\label{eq:Markov}
\end{equation}

On the other hand, because the algorithm errs with probability at most
$\epsilon$, we have for all $y$,
$\|P_y \ket{\phi^y_T}\|^2 \geq 1 -\epsilon, \label{goodcomponent}$, and
since $\|P_y \ket{\phi^y_T}\|^2 + \|(I-P_y) \ket{\phi^y_T}\|^2=1$, we
conclude that
\begin{equation}
\|(I-P_y) \ket{\phi^y_T}\| \leq \sqrt{\epsilon}. 
\label{badcomponent}
\end{equation}
By combining (\ref{eq:Markov}) and (\ref{badcomponent}) we conclude
that for all but an $N^{-\frac{1}{4}}$ fraction of the $y$'s in $[N]$,
\[ |\braket{\phi_T}{\phi_T^y}| 
\leq |\braket{\phi_T^y}{P_y|\phi_T}| + |\braket{\phi_T}{(I-P_y)|\phi_T^y}|
\leq \sqrt{\epsilon} + N^{-\frac{1}{4}}.\]
Let
$\theta_y \defeq \arccos\braket{\phi_T}{\phi_T^y}$. We claim that
for all but an $N^{-\frac{1}{4}}$ fraction of the $y$'s in $[N]$,
\begin{equation}
\theta_y \geq
\frac{\pi}{2}\left(1-O\left(\epsilon^{\frac{1}{2}}+
N^{-\frac{1}{4}}\right)\right).
\label{eq:thetay}
\end{equation}
To justify this, note that since $\epsilon \leq \frac{1}{10}$ and
$N\geq 100$, we have $\sqrt{\epsilon} + N^{-\frac{1}{4}} \leq
\frac{1}{\sqrt{2}}$ (say), and $\theta_y \geq \frac{\pi}{4}$. For
$\theta_y \leq \theta \leq \frac{\pi}{2}$,
$\frac{\mathrm{d}\cos\theta}{\mathrm {d} \theta} = - \sin \theta \leq
- \frac{1}{\sqrt{2}}$. So,
\[ \left(\frac{\pi}{2} - \theta_y\right) \times -\frac{1}{\sqrt{2}} 
\geq \cos \frac{\pi}{2} - \cos \theta_y
\geq -\left(\sqrt{\epsilon} + N^{-\frac{1}{4}}\right).\]
Our claim (\ref{eq:thetay}) follows from this.

Finally, since $\theta_y\geq 0$ for all $y$, we have
\[ \sum_y \theta_y = \sqrt{N}\left(1-N^{-\frac{1}{4}}\right)\frac{\pi}{2} \left(1-O\left(
\sqrt{\epsilon}+N^{-\frac{1}{4}}\right)\right) =
\sqrt{N}{\frac{\pi}{2}} \left(1-O\left(
\sqrt{\epsilon}+N^{-\frac{1}{4}}\right)\right).\]
\unskip\qed

\paragraph{Proof of Lemma~\ref{lm:smallpairwisedistance}:}
It follows from elementary trigonometry that for unit vectors $v$ and $w$,
\[\theta(v,w) = 2 \arcsin (\frac{1}{2} \min\{\| v-w \|, \|v + w\|)
 \leq 2 \arcsin (\frac{1}{2} \| v-w \|).
\]
On the other hand, since the oracle $O_y$ differs from the identity
transformation only for basis vectors in the subspace $\ket{y} \tensor
V$, one can show that (see, e.g., \cite{Zalka} for a similar
derivation)
\[ \| \phi_T^{y,i-1} - \phi_T^{y,i}\| \leq 2 \| P_y \ket{\phi_{T-i}}\|.\]
Our lemma follows from these two observations. \qed

\paragraph{Proof of Lemma~\ref{lm:equalprobability}:}
We will show that $\sum_y \arcsin \sqrt{p_{i,y}}$ achieves its maximum
when each $p_{i,y}$ is $\frac{1}{N}$. For, let $\sum_y \arcsin
\sqrt{p_{i,y}}$ take its maximum value at $(p_{i,y}: y \in [N])$. Now,
the function $\arcsin \sqrt{x}$ is concave in the interval
$[0,\frac{1}{2}]$ (for its second derivative is
$-\frac{1}{4}(1-2x)(x-x^2)^{\frac{3}{2}}$). Thus, our claim will
follow from Jensen's inequality if we can show that each $p_{i,y} \in
[0,\frac{1}{2}]$.  To show this, suppose for contradiction that for
some (necessarily unique) $\hy$, we have $p_{i,\hy} >
\frac{1}{2}$. Then,
\[
 \sum_y \arcsin \sqrt{p_{i,y}}  \leq (N-1) \arcsin \sqrt{\frac{1-p_{i,\hy}}{N-1}} 
 + \arcsin\sqrt{p_{i,\hy}}
\leq (1+o(1)) \sqrt{\frac{N-1}{2}} + \frac{\pi}{3}.
\]
But for large $N$, we have $N \arcsin \frac{1}{\sqrt{N}} \geq
\sqrt{N}$ is larger than this quantity. This contradiction implies
that our assumption about the existence of $\hy$ was incorrect. It
follows that
\[ \sum_y \arcsin \sqrt{p_{i,y}} \leq N \arcsin \frac{1}{\sqrt{N}} 
\leq \sqrt{N} \left(1+O\left(\frac{1}{N}\right)\right),\] where for
the first inequality we used the concavity of the function $\arcsin
\sqrt{x}$ for $x \in [0,\frac{1}{2}]$ to apply Jensen's inequality.
\qed

\end{document}